\documentclass[a4paper,12pt]{article}
    
\usepackage{abstract}

\usepackage{titlesec}
\titleformat{\section}{\large\bfseries}{\thesection}{1em}{}
\usepackage{authblk}

\usepackage[bottom]{footmisc}
    
\usepackage{fancyhdr}
\fancyheadoffset{0cm}
\providecommand{\keywords}[1]{\noindent\textbf{Keywords:} #1}

\usepackage{amssymb}
\usepackage{amsthm}

\usepackage{microtype}
\usepackage[utf8]{inputenc}
\usepackage[T1]{fontenc}
\usepackage{newtxtext}
\usepackage{newtxmath}
\usepackage{setspace}
\usepackage{lineno,xcolor}
\usepackage[top=25.4mm, bottom=25.4mm, left=25.4mm, right=25.4mm]{geometry}
    
\usepackage{graphicx}
\usepackage{float} 
    
\usepackage[hidelinks]{hyperref}

\usepackage{acronym}
\usepackage{upgreek}
\usepackage{booktabs}
\usepackage{cite}
\usepackage[font=small]{caption}

\newacro{AFM}{Atomic Force Microscopy}
\newacro{CL}{Cathodoluminescence}
\newacro{FIB}{Focused Ion Beam}
\newacro{PMT}{Photomultiplier Tube}
\newacro{PCB}{Printed Circuit Board}
\newacro{PGMEA}{Propylene Glycol Monomethyl Ether Acetate}
\newacro{QTF}{Quartz Tuning Fork}
\newacro{SE}{Secondary Electron}
\newacro{SEM}{Scanning Electron Microscopy}
\newacro{SNOM}{Scanning Near-field Optical Microscopy}
\newacro{SPM}{Scanning Probe Microscopy}
\newacro{TPP}{Two-Photon Polymerization}
\newacro{IBSD}{Ion Beam Sputtering Deposition}

\begin{document}

  \title{\vspace{-15mm}\LARGE\selectfont\textbf{
  A plug-and-play platform for self-aligning 3D-printed structures on optical fibers
  }}

  \author[1,2,$\dagger$]{\large Lukáš Zezulka}
  \author[1,2,$\dagger$, *]{\large Zdeněk Nováček}
  \author[1,2]{\large Martin Konečný}
  \author[3,4]{\large Milan Matějka}
  \author[4]{\large Štěpán Černý}
  \author[1,2]{\large Katarína Rovenská}
  \author[1,2]{\large Jiří Spousta}
  \author[1,2]{\large Tomáš Šikola}
  
  \affil[1]{\normalsize Brno University of Technology, Faculty of Mechanical Engineering, Institute of Physical Engineering, Technická 2, 61669 Brno, Czech Republic}
  \affil[2]{\normalsize Brno University of Technology, Central European Institute of Technology, Purkyňova 656/123, 61200 Brno, Czech Republic}
  \affil[3]{\normalsize Institute of Scientific Instruments of the Czech Academy of Sciences, Královopolská 147, 61264 Brno, Czech Republic}
  \affil[4]{\normalsize IQS nano s.r.o., Hlavní 130, 25068, Husinec, Czech Republic
  \vspace{15pt}}
  \affil[$\dagger$]{\normalsize These authors contributed equally to this work.}
  \affil[*]{\normalsize Corresponding author. E-mail: novacek@fme.vutbr.cz}

  \renewcommand\Authands{ and }

\date{}

\maketitle 
\vspace{-5mm}
\begin{abstract} 
\textit{
Two-photon polymerization 3D printing advances optical fibers into complex lab-on-fiber devices. However, widespread adoption remains bottlenecked by the inherent inefficiencies of sequential, direct-on-fiber printing. To address this, a custom-designed modular platform and a corresponding fiber-decoupled fabrication methodology on planar substrates are presented. This platform features a compliant, spring-like socket that enables passive self-alignment and reversible fiber insertion, paired with a customizable frontal device. The shift to fabrication on planar substrates enables parallel batch processing and allows the substrate to serve as a functional magazine for device mounting. To demonstrate this approach, modular platforms featuring a functional tip are printed, sharpened via post-process plasma etching, metallized, and equipped with an optical aperture using focused ion beam milling, while remaining on the planar substrate. Subsequently, these are used as exchangeable scanning probe microscopy tips mounted in situ within a scanning electron microscope. Simultaneous three-channel correlative imaging (atomic force microscopy topography, secondary electron, and cathodoluminescence) is used to validate the optomechanical stability of the design. Ultimately, by decoupling the delicate optical device's lifespan from the optical fiber, this architecture transforms lab-on-fiber tools into scalable, seamlessly replaceable consumables, offering a highly adaptable pathway for broader applications in photonics, sensing, and microscopy.
}
\end{abstract}

\setlength\parindent{.45in} \keywords{two-photon polymerization, lab-on-fiber, self-alignment, fiber-decoupled fabrication, optical fiber connector, scanning probe microscopy}

\section{Introduction}
Optical fibers have evolved from passive light guides into versatile platforms for integration of multifunctional devices \cite{pendaoOptical2022}. In particular, lab-on-fiber technologies have transformed the fiber tip into a compact site for integration of devices tailored to specific applications \cite{piscoLabOnFiber2020, xiongMultifunctional2020}.

Within this progress, femtosecond-laser \ac{TPP} has proven to be a powerful resource for directly fabricating complex micro-optomechanical components on optical fiber facets \cite{bianFemtosecond2024}. \ac{TPP} provides freeform three-dimensional design capabilities, sub-micrometer fabrication resolution, and precise alignment with the fiber core. This overcomes the geometric limitations of conventional fiber-end optics and enables comprehensive control of the emerging optical field \cite{arriaga-davilaSingle2025,gissiblSubmicrometre2016, gonzalez-hernandezMicroOptics2023, wangTwophoton2024}. These capabilities have been utilized in diverse application-specific fiber-tip devices, including beam-shaping elements, sensors, mechanical grippers, and \ac{SPM} probes \cite{gissiblSubmicrometre2016, lai3D2024, bertonciniPolarization2018, caoTwophoton2022, chenTwophoton2024, kimFiberOptic2020, liUltrathin2020, bauerSidelooking2025, litactile2025, zouFibertip2021, alsharifDesign2018}. Within the domain of scanning probes, 3D-printed \ac{SPM} architectures have demonstrated the ability to leverage complex microstructures to tune resonance spectra and tip-sample impact \cite{goringTailored2016, sun3Dprinted2020}. Furthermore, light-guiding \ac{SNOM} tips have been introduced for asymmetric nanofocusing \cite{sun3D2018}, high-numerical-aperture mapping, and ultra-sharp tip curvature radii down to 15\,nm \cite{jungFabrication2013, zhouSharp2023, jungFabricationof2013}.

Despite these advances, fabricating devices directly on optical fibers introduces severe logistical and physical bottlenecks inherent to the quasi-one-dimensional nature of the fiber substrate. Each fiber must be handled individually, significantly prolonging the manipulation time required for mounting and dismounting during fabrication. This inefficiency is especially pronounced when managing long, bundled fibers, and propagates through all subsequent post-processing steps, including thin-film coating, plasma etching, and pyrolysis. Furthermore, monolithic integration couples the structural integrity of the optical fiber directly to the printed device; damage to either component renders the entire assembly nonfunctional. This coupled risk is especially adverse for short-lifetime or intentionally consumable devices, ultimately limiting the scalability and adoption of 3D-printed on-fiber technologies.

These limitations can be overcome by decoupling device fabrication from the optical fiber and instead printing structures on a planar substrate for subsequent mounting. This shifts manufacturing to a planar wafer-scale approach, facilitating fabrication and processing of structures in batches. A planar substrate streamlines typical post-processing workflows—including curing \cite{purtovImproved2018}, plasma etching \cite{powerBistable2023}, pyrolysis \cite{seniutinas1002018}, thin-film deposition \cite{astrauskyteAntiReflective2023}, and surface functionalization \cite{zhangFacile2021}—while facilitating a batch-level visual inspection to isolate the highest-performing structures \cite{vanmolFabrication2024, siegleComplex2023}. Decoupled fabrication is highly advantageous for rapid prototyping and consumable applications subject to rapid wear or cross-contamination \cite{rovatiConstruction2011, handisposable2025}. However, this paradigm shift is only viable if the subsequent fiber-integration procedure is sufficiently robust and time-efficient \cite{djogoSelfAligning2024, xu3Dprinted2023}.

The foundation of decoupled integration has been demonstrated using planar-printed sockets for the passive insertion and self-alignment of micro-optics onto fibers \cite{s3DPrintedFiber2020}. While this prior architecture provides a proof-of-concept, it is limited by a large form factor; accommodating a standard 125$\,\upmu\mathrm{m}$ diameter fiber required a socket 470$\,\upmu\mathrm{m}$ in diameter and 500$\,\upmu\mathrm{m}$ in length. Such dimensions restrict practical application in domains that require slender, low-profile, or minimally invasive probes, such as \ac{SPM}. To address this gap, we introduce a compact, modular platform that maintains the scalability of decoupled planar fabrication while minimizing the mechanical footprint. The viability of this platform is validated by a high-stress proof-of-concept application that demonstrates robust optomechanical stability during correlative microscopy consisting of \ac{AFM}, \ac{SEM}, and \ac{CL}.

\section{Concept and design of the platform}
\begin{figure}[htbp]
\centering\includegraphics{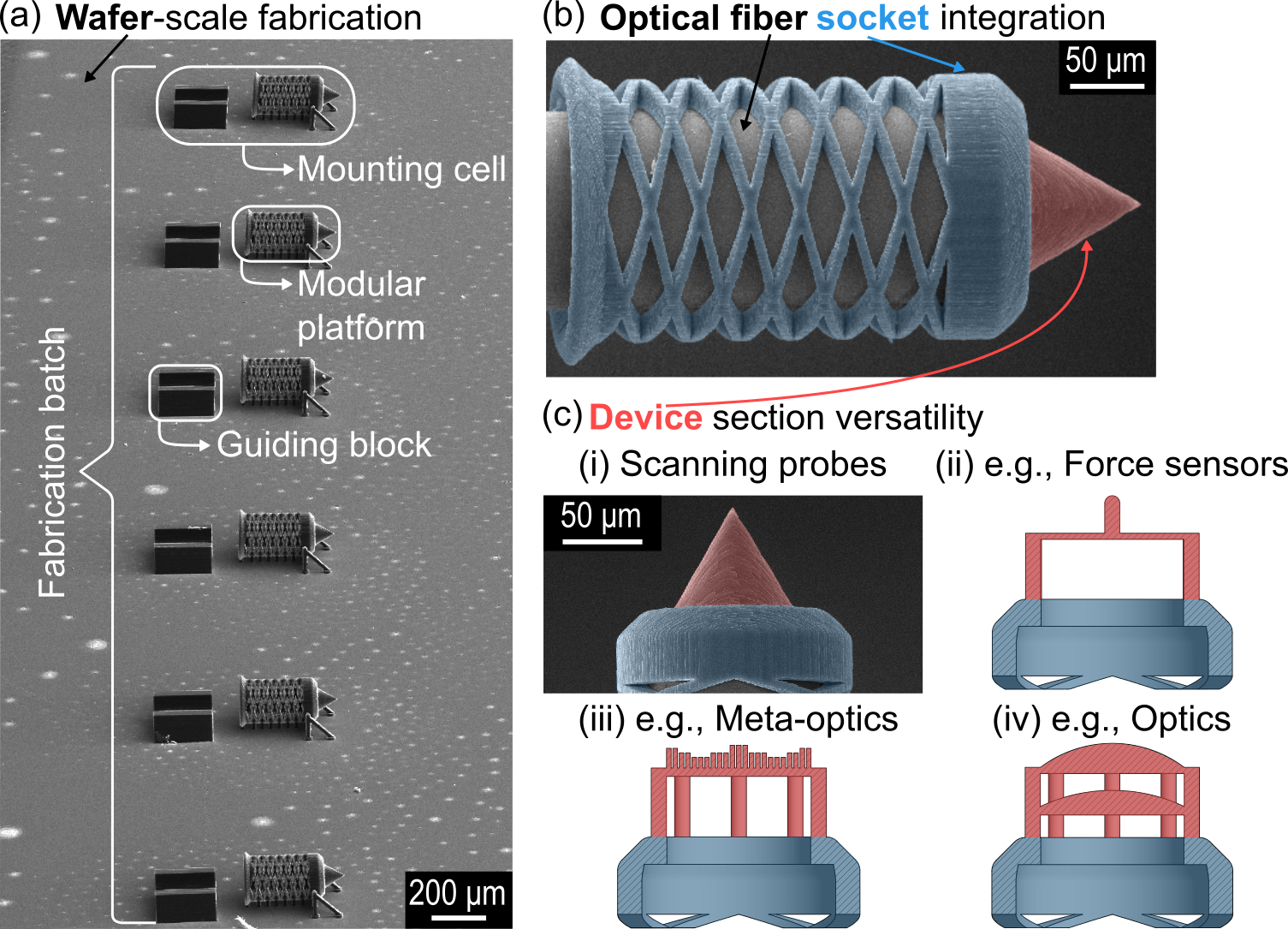}
\caption{\ac{TPP} 3D-printed modular platforms enabling fiber-decoupled, wafer-scale fabrication of functional devices. 
(a) \ac{SEM} image of a fabrication batch consisting of multiple mounting cells printed on a planar substrate. 
(b) Colored \ac{SEM} image of the modular platform highlighting the socket section (blue) for optical fiber integration and the device section (red) in the form of an \ac{SPM} tip. 
(c) Proposed versatility of the device section; (i) colored \ac{SEM} image, (ii--iv) illustrations.
}
\label{fig1}
\end{figure}

Here, we propose a modular platform compatible with serial wafer-scale fabrication (Figure~\ref{fig1}a), comprising a socket for passive optical fiber integration and a customizable device section (Figure~\ref{fig1}b). Key design parameters prioritize an efficient, reversible fiber-mounting process, mechanical stability, and effective light coupling across the fiber-device interface. This architecture is inherently versatile and can support a wide array of \ac{TPP} 3D-printed optical fiber devices (Figure~\ref{fig1}c). For validation, correlative \ac{AFM}, \ac{SEM}, and \ac{CL} imaging utilizing an integrated \ac{SNOM} tip as the sensing device was selected as a demanding proof-of-concept application. Consequently, this specific modular platform will be referred to as the \textbf{exchangeable \ac{SPM} tip} throughout the application-related sections of this work. This high-stress application explicitly tests mechanical stability under shear-force scanning, optical coupling, and structural integrity inside a high-vacuum environment. Rather than optimizing the fundamental optics of the \ac{SNOM} tip, which has already been established in the literature \cite{jungFabrication2013, jungFabricationof2013}, the tip is used strictly as an optomechanical test to validate the performance and parallel post-processing advantages of the presented modular platform.

\begin{figure}[htbp]
\centering\includegraphics{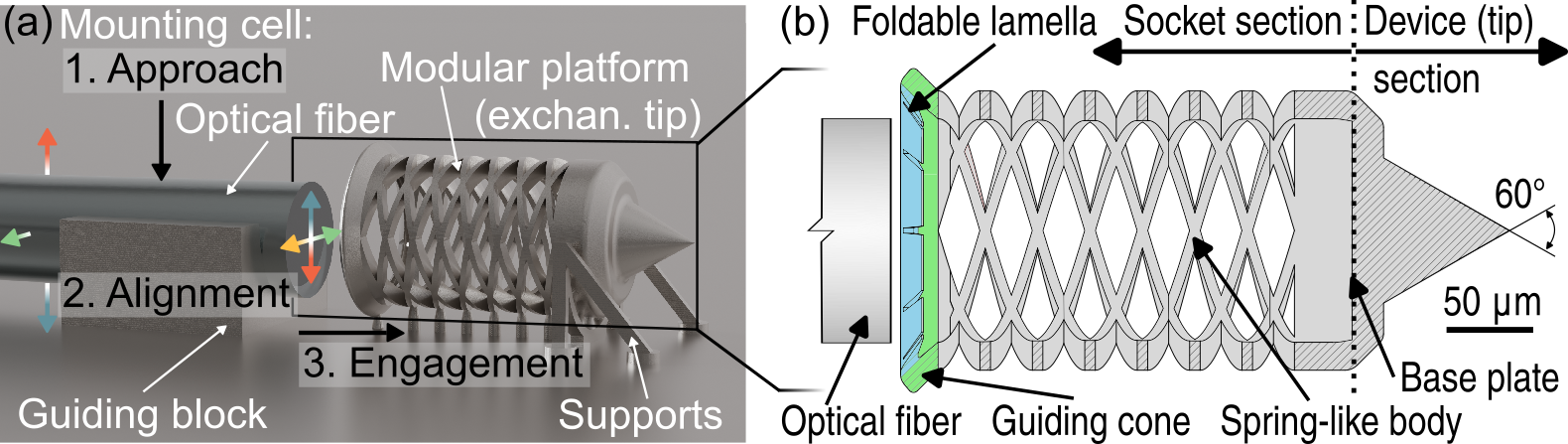}
\caption{Architecture and mounting methodology of the \ac{TPP} 3D-printed modular platform. 
(a) 3D rendering of a single mounting cell on the planar substrate, comprising the modular platform (acting as the exchangeable \ac{SPM} tip), auxiliary supports, and a guiding block. The sequence illustrates the fiber approach, alignment (highlighted by the colored directional arrows), and engagement. During alignment, the guiding block serves as a solid contact point against which the optical fiber is levered. 
(b) Cross-sectional design detailing the compliant socket section for fiber integration, the solid internal base plate, and the functional device section (configured here as an \ac{SPM} tip).
}
\label{fig2}
\end{figure}

The modular platform body consists of the functional device section and a structurally compliant socket section (Figure~\ref{fig2}b). The socket serves as a self-locking sleeve that guides and secures the optical fiber. The mechanical design dictates a specific mounting sequence: at the rear of the socket, a guiding cone enables smooth optical fiber insertion. The inner wall of this cone incorporates foldable lamellae featuring a nominal interference fit of 1$\,\upmu\mathrm{m}$ at the center that tapers toward the edges. These lamellae permit smooth forward fiber insertion while strongly resisting retraction. During insertion, friction against the lamellae longitudinally compresses the central spring-like body. This compression induces a slight radial expansion of the spring structure, facilitating the continuous advancement of the fiber. The spring body transitions into a rigid, solid-walled front section where the fiber facet is seated flush against the optical base plate. This compliant geometry accommodates common \ac{TPP} post-fabrication shrinkage and provides tolerance to pre-mounting misalignment while decreasing the overall polymerized volume compared to a solid-walled cylinder, thereby significantly reducing the printing time. 

The frontal segment of the platform houses the device section, which in this paper takes the form of an \ac{SPM} tip for validation (Figure~\ref{fig1}c(i)). The tip is structured as a cone with a $60^{\circ}$ apex angle; this geometry maximizes optical throughput toward the terminal aperture \cite{novotnyPrinciplesNanoOptics2012} while providing sufficient clearance for unobstructed electron-beam access during correlative \ac{SEM} imaging.

To fully leverage the aforementioned wafer-scale fabrication capabilities, the mounting cell design embeds auxiliary structures so that the planar substrate can be used as an exchangeable tip magazine (Figure~\ref{fig2}a). Each platform is fixed to the substrate via supporting struts equipped with predefined breaking points, allowing a clean mechanical release upon successful fiber engagement. Additionally, a guiding block is printed directly behind each socket. This block pre-aligns the approaching optical fiber with the socket's longitudinal axis, significantly increasing the reliability and speed of the mounting procedure.

\section{Fabrication and post-processing}
Fabrication was performed via \ac{TPP} using an IQnano3D system (IQS nano) equipped with a 515\,nm femtosecond laser source (170\,fs pulse duration, 80\,MHz repetition rate, 200\,mW average power) and a $100\times$ microscope immersion objective (NA = 1.25, Olympus). The mounting cells were fabricated from ResIQ Hybrid (IQS nano), an ORMOCER-type organic-inorganic hybrid photopolymer containing an organically modified silicon-oxygen-based inorganic network integrated within a cross-linked polymer matrix. This material was selected for its compatibility with high-resolution 3D-printing, its mechanical rigidity (Young's modulus $\approx 1\,\mathrm{GPa}$), thermal stability (up to $270\,^\circ\mathrm{C}$), and optical properties (refractive index $n \approx 1.5$) \cite{IQSresiq}.

\begin{figure}[htbp]
\centering\includegraphics{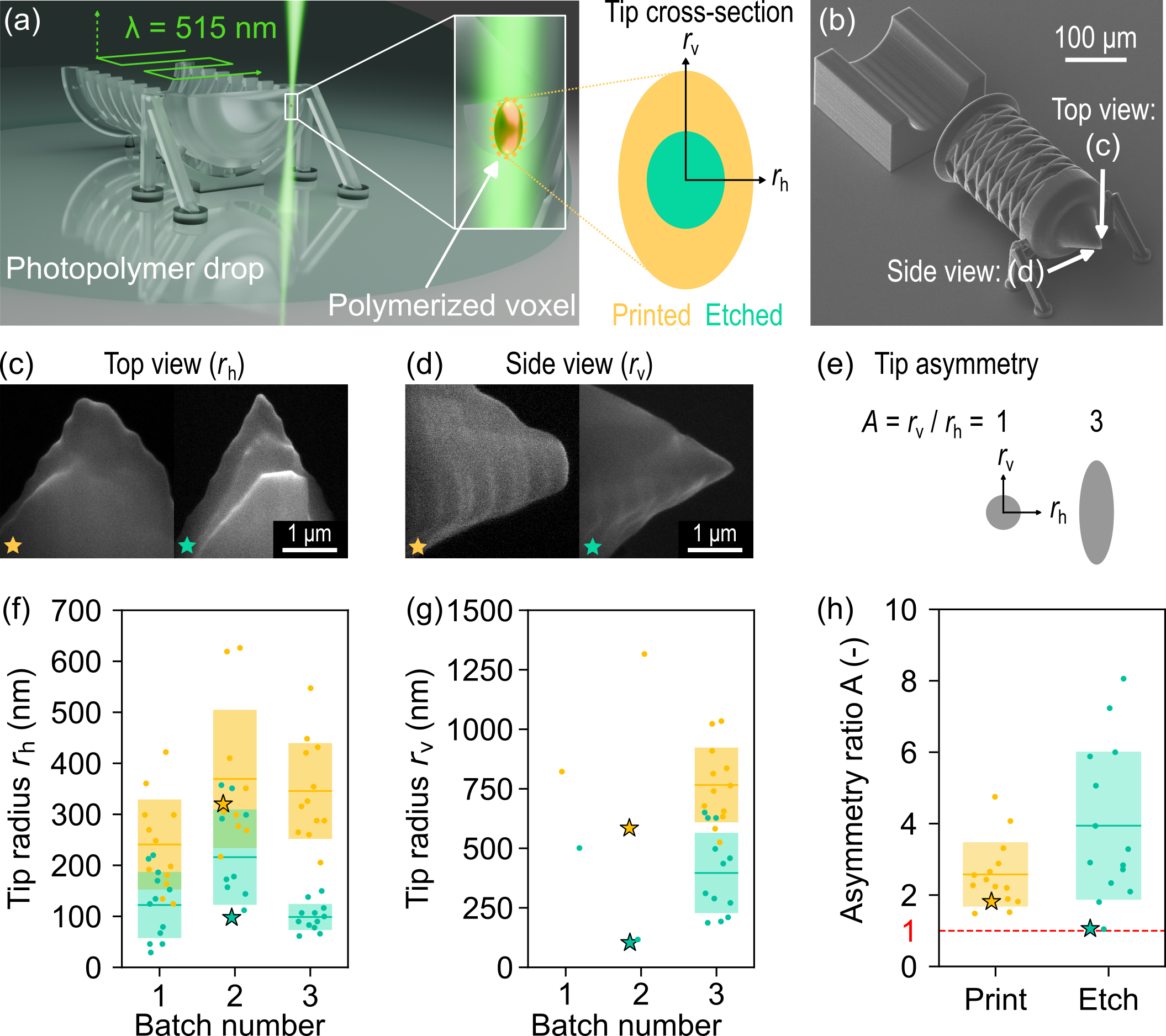}
\caption{Fabrication and characterization of the \ac{TPP} 3D-printed exchangeable \ac{SPM} tips and the effect of subsequent etching on the tip sharpness. 
(a) Schematic representation of the \ac{TPP} layer-by-layer 3D-printing process in a polymer drop. The insets illustrate a polymerized voxel and a cross-sectional view defining the horizontal ($r_\mathrm{h}$) and vertical ($r_\mathrm{v}$) tip radii before (yellow, printed) and after (green, etched) the etching process. 
(b) \ac{SEM} image of a single \ac{TPP} 3D-printed mounting cell, with the arrows indicating the viewing angles for the two subsequent panels. 
(c) Top and 
(d) side view \ac{SEM} images comparing the tip morphologies of as-printed (left) and etched (right) structures. The yellow and green stars denote the representative symmetrical tip for the as-printed and etched states, respectively. 
(e) Schematic defining the tip asymmetry ratio as $A = r_\mathrm{v} / r_\mathrm{h}$. 
Statistical distribution of (f) the horizontal tip radius $r_\mathrm{h}$, 
and of (g) the vertical tip radius $r_\mathrm{v}$.
(h) The calculated asymmetry ratio $A$ across three different fabrication batches. While each batch nominally consists of 12 printed samples (with two tips omitted in the second batch due to a printing error), only a subset of tips (1, 2, and 12 tips for the first, second, and third batches, respectively) were fully imaged from the side due to the gradual development of the imaging methodology. The data are compared for as-printed (yellow) versus etched (green) tips and are presented as the average with standard deviation bounds. The star symbols have the same meaning as in (c). The red dashed line in (h) indicates perfect symmetry ($A = 1$).}
\label{fig3}
\end{figure}

To enable the wafer-scale, fiber-decoupled paradigm, the modular platforms were printed in a horizontal orientation on a planar silicon substrate with the longitudinal axis of the modular platform and the integrated tip oriented approximately parallel to the substrate surface. Consequently, the tip axis was oriented orthogonal to the optical axis of the printing beam (Figure~\ref{fig3}a). 

The mounting cell was imported as a three-dimensional STL model occupying an area of approximately $675\,\upmu\mathrm{m}\times240\,\upmu\mathrm{m}$ with a total height of $197.5\,\upmu\mathrm{m}$. This model was processed using the proprietary IQ Data software with the IQ Slicer module, which converted the model into a sequence of fabrication layers and corresponding exposure points. The STL model was discretized into 792 layers with a vertical layer spacing of $0.25\,\upmu\mathrm{m}$. The lateral voxel-to-voxel spacing in the horizontal plane was set to $0.25\,\upmu\mathrm{m}$. 

Before printing, the substrate-photoresist interface was automatically detected at four positions corresponding to the corners of the print area. These measurements were used for automatic leveling and compensation of substrate tilt or local height variations. To ensure reliable attachment of the printed structures to the substrate, the starting layer was embedded by $0.25\,\upmu\mathrm{m}$ below the detected substrate-resin interface. The three batches presented in this study were fabricated using an optimized exposure time of $0.5\,\upmu\mathrm{s}$. Under these conditions, the total printing duration for one complete mounting cell was approximately 25 minutes.

Following \ac{TPP} fabrication, the samples were developed in \ac{PGMEA} for 5\,min and then put into a second clean \ac{PGMEA} bath for an additional 5\,min. Subsequently, the samples were transferred into an isopropanol bath for 5\,min, rinsed with fresh isopropanol spray, and finally dried using clean dry air or nitrogen. This development sequence was used to remove unpolymerized resin while minimizing contamination and residue formation on the printed structures.

Three batches of 12 exchangeable tips each were printed using this procedure; however, two tips were omitted from the second batch due to a queue error. The horizontal orientation of the exchangeable tip on the substrate, and consequently the orthogonal orientation of the tip to the laser beam (Figure~\ref{fig3}a), causes a structural asymmetry at the very apex of the tip. This is an inherent property of the \ac{TPP} voxel, which exhibits an elongated ellipsoidal geometry along the optical axis, causing the achievable axial feature size to be fundamentally larger than the lateral one \cite{ohalloranTwoPhoton2023}.

To characterize the fabricated tips, two principal radii were defined and measured: the radius in the horizontal plane ($r_\mathrm{h}$ -- measured in top view) and in the vertical plane ($r_\mathrm{v}$ -- measured in side view), as detailed in Figure~\ref{fig3}a. These dimensions were extracted from \ac{SEM} micrographs (Figure~\ref{fig3}c,d) by manually fitting a circle to the tip apex. For structures having a flattened plateau, the fit was adapted across the degraded geometry to estimate the effective macroscopic curvature. To strictly preserve the native tip morphology, all \ac{SEM} characterization was performed on pristine samples without charge dissipation coatings.

The analysis of the as-printed structures yielded average horizontal radii of $r_\mathrm{h} = 241\pm88\,\mathrm{nm}$, $369\pm186\,\mathrm{nm}$, and $346\pm94\,\mathrm{nm}$ across the three consecutive batches (Figure~\ref{fig3}f). Due to the progressive refinement of the imaging methodology, comprehensive side-view characterization was restricted to the third batch, which exhibited an average vertical radius of $r_\mathrm{v} = 766\pm157\,\mathrm{nm}$ (Figure~\ref{fig3}g). Incorporating the limited side-view data from the first (one tip) and second (two tips) batches, the global average tip asymmetry—defined as the ratio $A=r_\mathrm{v}/r_\mathrm{h}$ (Figure~\ref{fig3}e)—was calculated to be $A = 2.5\pm0.9$ (Figure~\ref{fig3}h).

The limitation of spatial resolution inherent to the orientation of the tip relative to the printing laser is especially evident when compared to structures printed directly onto optical fibers parallel to the optical axis, which featured curvature radii of $125\,\mathrm{nm}$ \cite{jungFabricationof2013}. Nevertheless, the optimal as-printed tip reported here achieved a comparable horizontal radius of $r_\mathrm{h} = 124\,\mathrm{nm}$.

To surpass the resolution limits of \ac{TPP} 3D-printing, post-fabrication plasma etching was employed. Prior studies have shown significant sharpening of fiber-printed tips using oxygen plasma on SCR500 resin, reducing the radii from $125\,\mathrm{nm}$ to $15\,\mathrm{nm}$ \cite{jungFabrication2013}. The presented planar substrate architecture, which carries the platforms, seamlessly facilitates the parallel application of a similar etching protocol.

The etching parameters were set to a forward power of $100\,\mathrm{W}$, a working pressure of $100\,\mathrm{mTorr}$, and a processing time of $300\,\mathrm{s}$. A flux of $30\,\mathrm{sccm}$ O$_2$ and $30\,\mathrm{sccm}$ SF$_6$ was utilized. This specific ratio was critical for uniformly etching both the organic and inorganic components of the polymerized matrix. Although pure O$_2$ plasma can enhance tip sharpness, it aggressively and preferentially removes organic compounds, resulting in severe morphological bulging (Figure~S7 in Supplement~1).

Following the plasma treatment, the average horizontal tip radii across the three batches were reduced to $r_\mathrm{h} = 122\pm65\,\mathrm{nm}$, $216\pm94\,\mathrm{nm}$, and $99\pm26\,\mathrm{nm}$ (Figure~\ref{fig3}f). Concurrently, the average vertical radius in the third batch decreased to $r_\mathrm{v} = 397\pm169\,\mathrm{nm}$ (Figure~\ref{fig3}g). The minimum recorded horizontal radius for a single etched tip was $r_\mathrm{h} = 29\,\mathrm{nm}$. Avoiding charge dissipation coatings for the \ac{SEM} characterization enabled direct morphological comparisons of each tip before and after etching (Figure~\ref{fig3}c,d and Figures~S1--S6 in Supplement~1). For the third batch, the etching protocol reduced the horizontal and vertical radii by average factors of 3.5 and 1.9, respectively. This anisotropic etch rate inherently increased the average structural asymmetry to $A = r_\mathrm{v}/r_\mathrm{h} = 3.8\pm2.1$ (Figure~\ref{fig3}h), although two exemplary structures achieved a near-perfect symmetry ($A\approx1$). The yield of symmetrical tips could be increased by rigorously optimizing parameters to establish highly anisotropic etching regimes that preferentially target the larger vertical dimension \cite{rasmussenSU82013}.

Despite these geometric variations, the achieved dimensions meet the requirements for aperture-type \ac{SNOM} probes, which typically utilize subwavelength apertures ranging from tens to hundreds of nanometers. Because the terminal metal-coated aperture forms the primary optical interface, it fundamentally dictates the local field confinement and coupling behavior. Consequently, the structural asymmetry of the dielectric support is not expected to compromise optical performance, provided the terminal aperture remains well-defined and sufficiently symmetric \cite{novotnyPrinciplesNanoOptics2012}.

\begin{figure}[htbp]
\centering\includegraphics{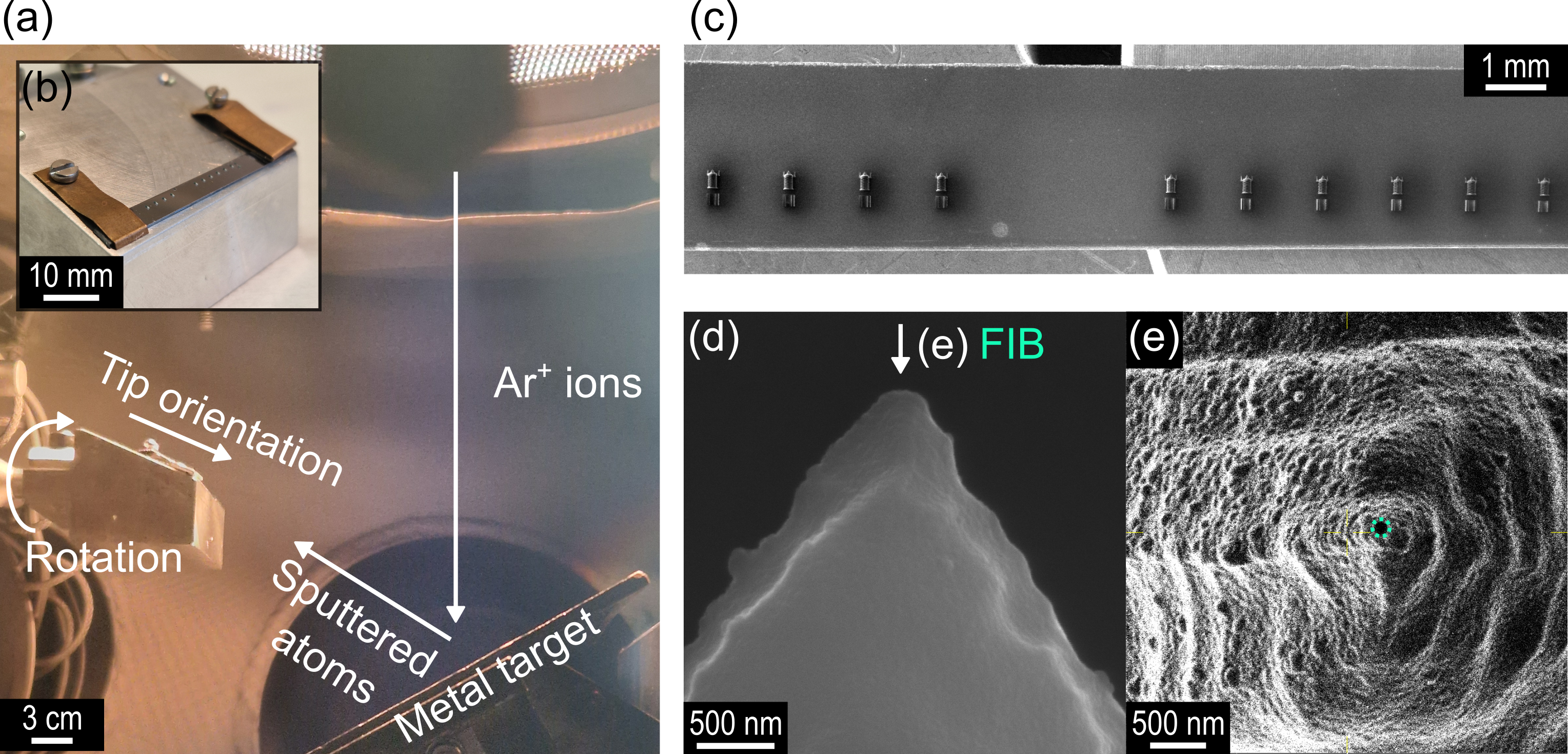}
\caption{Batch-level metallization and aperture fabrication of the \ac{TPP} 3D-printed exchangeable \ac{SPM} tips. 
(a) Camera image of the \ac{IBSD} vacuum chamber used for the deposition of a $5\,\mathrm{nm}$ Ti adhesion layer followed by a $150\,\mathrm{nm}$ Al film. 
(b) Detail of the sample holder, illustrating the positioning of the printed exchangeable tips towards the edge; the tips are oriented directly toward the sputtered atoms and rotated during deposition to minimize shadowing. 
(c) \ac{SEM} image of the coated mounting cells from the second fabrication batch (with two tips omitted due to a previous printing queue error). 
(d) Top-view \ac{SEM} detail of a single tip, showing the morphology of the metallic coating over the etched apex. 
(e) Front-view \ac{FIB} \ac{SE} image demonstrating a $100\,\mathrm{nm}$ wide aperture milled with Ga$^+$ \ac{FIB} at the apex.}
\label{fig4}
\end{figure}

Following plasma etching, the whole fabricated batch underwent simultaneous surface metallization without removal from the planar substrate. An \ac{IBSD} system (Figure~\ref{fig4}a) was used to deposit a $5\,\mathrm{nm}$ titanium adhesion layer followed by a $150\,\mathrm{nm}$ aluminum thin film. The process was performed using argon ions at an acceleration voltage of $600\,\mathrm{V}$, at a base pressure of $1.6 \times 10^{-6}\,\mathrm{mbar}$ and a working pressure of $3.1 \times 10^{-4}\,\mathrm{mbar}$. The deposition rates were approximately $0.3\,\text{\AA}\cdot\mathrm{s}^{-1}$ for Ti and $0.5\,\text{\AA}\cdot\mathrm{s}^{-1}$ for Al.

To minimize geometric shadowing, the substrate was mounted at a tilt relative to the atom flux, positioned near the front edge of the sample holder (Figure~\ref{fig4}b), and continuously rotated during the deposition. This configuration ensured sufficient metal coverage of the exposed exterior surfaces (Figure~\ref{fig4}c,d) while intentionally shadowing the socket's inner optical base plate, preserving its transparency for subsequent light coupling. Finally, a subwavelength optical aperture was fabricated at the tip apex by Ga$^+$ \ac{FIB} milling (Figure~\ref{fig4}e). While \ac{FIB} milling is inherently a sequential process, retaining the tips on the planar substrate enabled continuous serial milling across the entire batch within a single vacuum cycle, streamlining the fabrication workflow as well as minimizing handling risks.

\section{Fiber mounting}

While the presented modular platform was successfully tested directly on bare optical fibers, its versatility is demonstrated here by mounting it onto an optical fiber attached to a \ac{QTF} of an \ac{SPM} module compatible with high vacuum (LiteScope, NenoVision \cite{novotnaAFMinSEM2020}).  This \ac{SPM} configuration was chosen as a proof-of-concept scenario that requires mechanical rigidity against high-frequency shear forces, precise optical alignment for light collection, and vacuum-chamber compatibility.

When the fabrication batch is installed into the custom dual-sample configuration of the \ac{SEM}/\ac{SPM} setup, it effectively serves as a tip magazine. This modularity allows for the in situ mounting and replacement of the 3D-printed exchangeable \ac{SPM} tip directly from its stationary mounting cell onto a long-term installed high-performing \ac{QTF} probe. 

Although 3D-printed polymer tips have demonstrated sufficient durability even under demanding contact-mode scanning \cite{goringTailored2016}, they remain inherently more susceptible to mechanical wear than conventional probes. Consequently, this modularity enables the replacement of a tip upon eventual degradation or contamination without dismantling the whole optical fiber probe. Such an in situ exchange is facilitated by the aforementioned LiteScope \ac{SPM} module. At a $0^{\circ}$ \ac{SEM} stage tilt, the \ac{SEM} monitors as the \ac{SPM} probe accesses the horizontally mounted tip magazine. By tilting and rotating the \ac{SEM} stage, both the probe and tip apex can be continuously observed as they access the region of interest on the vertically mounted target sample.

\begin{figure}[htbp]
\centering\includegraphics{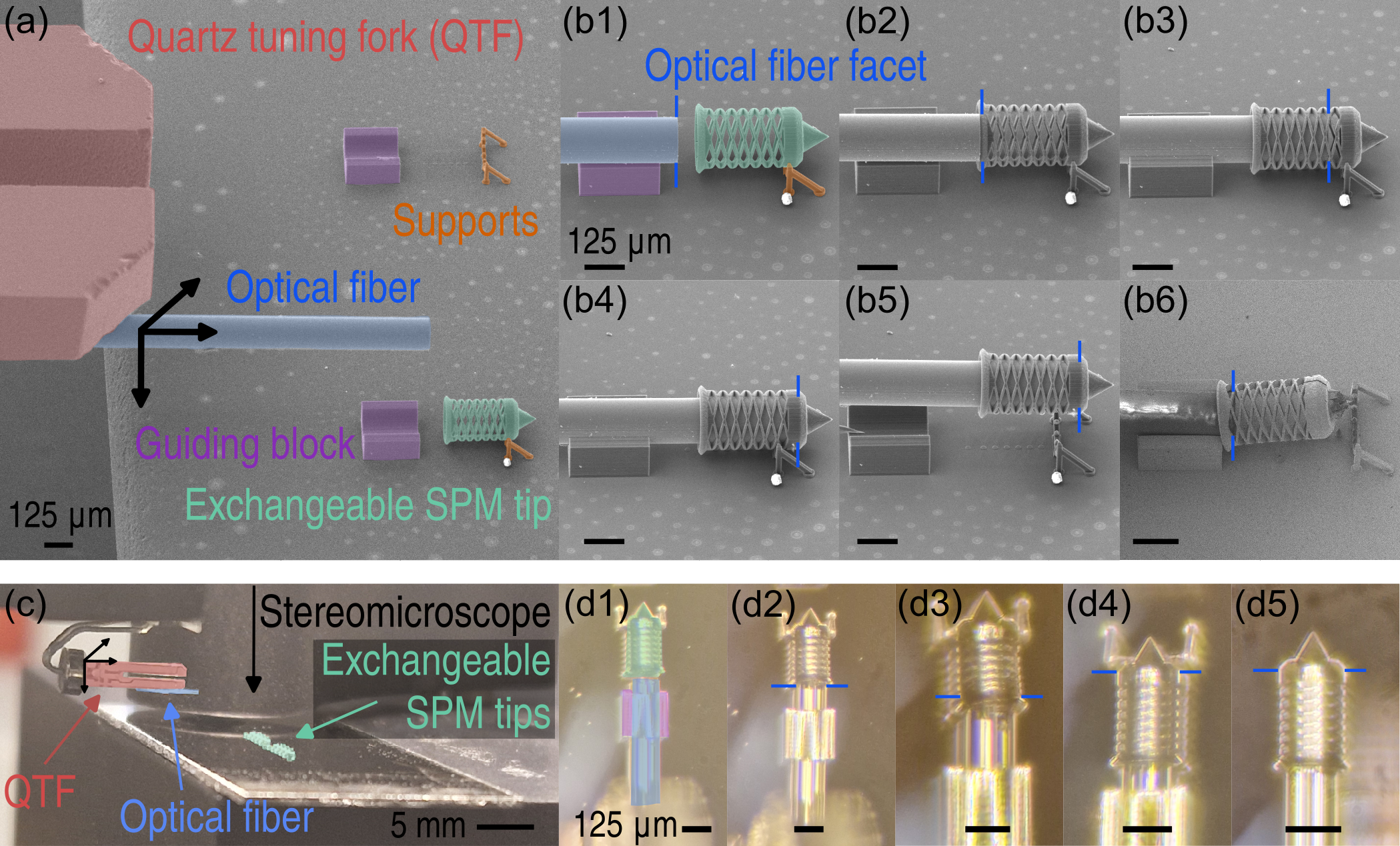}
\caption{High-precision in situ and manual benchtop mounting of the \ac{TPP} 3D-printed exchangeable \ac{SPM} tips. 
(a) False-colored \ac{SEM} overview of the mounting setup, showing an optical fiber (blue) attached to a \ac{QTF} (red) approaching a guiding block (purple) and an exchangeable \ac{SPM} tip (green). 
(b1--b5) \ac{SEM} image sequence of the mounting workflow. The fiber is aligned, inserted through the guiding block (b1), and pushed into the tip's spring-like socket (b2). The spring provides visual feedback for fine alignment (b3), and pushing forward a fully engaged fiber (b4) breaks it free from the supporting struts (b5). A continuous recording of the in situ \ac{SEM} mounting procedure is shown in Visualization 1.
(b6) \ac{SEM} image of the dismounting process, where retracting the fiber along the guiding block mechanically strips the used tip from the facet. 
(c) Photograph of a simplified benchtop mounting setup utilizing a stereomicroscope and manual micro-positioners. 
(d1--d5) Optical stereomicroscope image sequence demonstrating the successful tip mounting using the manual setup, facilitated by the socket's inherent mechanical self-alignment.}
\label{fig5}
\end{figure}

The high-precision in situ mounting workflow leverages the motorization of the \ac{SPM} and live visual feedback from the \ac{SEM} (Figure~\ref{fig5}a,~b1--b5). The fiber insertion procedure consists of three distinct steps:
\begin{enumerate}
\item \textbf{Approach and Alignment:} The optical fiber is driven toward the magazine and passed through a guiding block. This block mechanically corrects residual tilt and rotation, ensuring strict collinearity between the fiber and the selected tip's axis (Figure~\ref{fig5}b1).
\item \textbf{Engagement:} The fiber is advanced into the spring-like socket. The physical compression of the spring provides critical visual feedback, enabling immediate corrections (Figure~\ref{fig5}b2,~b3).
\item \textbf{Detachment:} Once the optical fiber facet seats against the inner base of the socket, a final forward motion fractures the predetermined breaking points on the supporting struts, leaving the \ac{SPM} tip securely mounted on the fiber (Figure~\ref{fig5}b4,~b5).
\end{enumerate}
Using this mounting procedure, a 95\,\% success rate was achieved.

Furthermore, the mounting mechanism is fully reversible, enabling the rapid detachment of worn or contaminated tips (Figure~\ref{fig5}b6). By retracting the optical fiber backward through the rigid guiding block, the used tip is mechanically stripped from the facet. This process leaves the optical fiber pristine and immediately ready for a subsequent mounting cycle without requiring probe removal or manual intervention. It is important to note that while the dismounting process leaves the optical fiber pristine for subsequent mounting cycles, the \ac{TPP} 3D-printed exchangeable tip itself is designed strictly as a single-use consumable. The physical engagement relies on the intentional deformation of the compliant structure and requires the supporting structures to secure the platform during mounting; thus, a dismounted platform cannot be reused.

To evaluate the platform's dependence on high-precision instrumentation, the mounting procedure was additionally tested using a manual benchtop setup. As shown in Figure~\ref{fig5}c and d1--d5, the fiber was mounted under a standard optical stereomicroscope using a manual 3-axis micro-positioner. Despite the mechanical self-alignment provided by the guiding block and spring-like socket, the procedure yielded a 70\,\% success rate. Experimental observations indicate that this reduced yield is primarily bottlenecked by the limited optical resolution of the stereomicroscope—failing to provide sufficient visual feedback during the critical spring-engagement phase—rather than by the motion-precision constraints of manual micro-positioning. Consequently, while manual benchtop mounting is physically feasible, achieving high reliability requires high-definition monitoring.

\section{Proof-of-concept: validation by SPM application}

\begin{figure}[htbp]
\centering\includegraphics{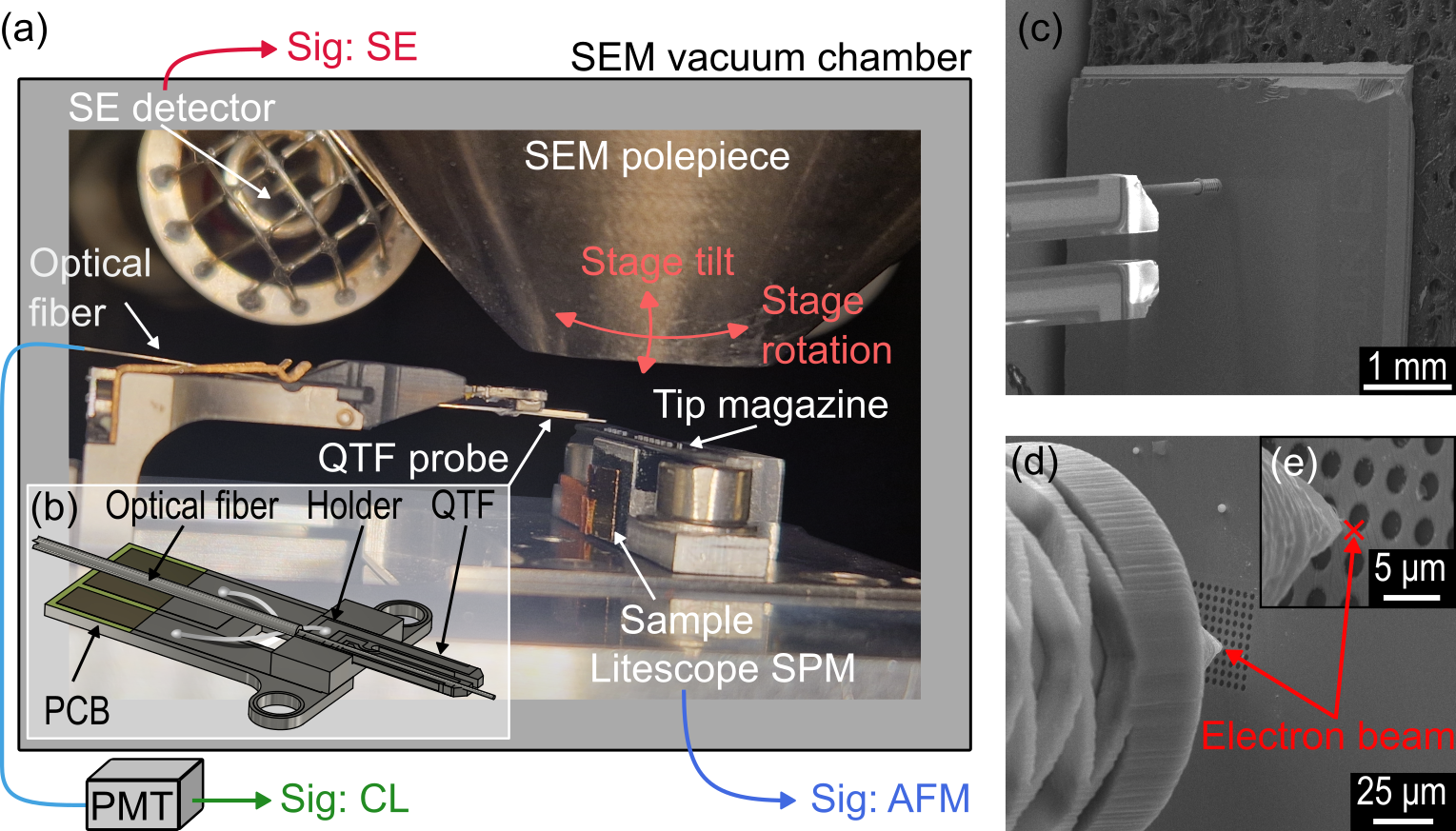}
\caption{Experimental setup and measurement configuration for correlative \ac{AFM}, \ac{SEM}, and \ac{CL} provided by an \ac{SNOM} tip. 
(a) Photograph of the LiteScope \ac{SPM} module integrated within an \ac{SEM} chamber. The setup adopts a custom dual-sample configuration, featuring a horizontally mounted tip magazine for the in situ tip exchange and a vertically oriented sample for measurement. The custom \ac{QTF} probe houses an integrated optical fiber enabling simultaneous extraction of the \ac{CL} signal, generated by the electron beam of the \ac{SEM}, and \ac{AFM} sensing of surface morphology. The \ac{CL} signal is detected by a \ac{PMT}.  
(b) Visualization of the custom-designed \ac{QTF} probe. 
(c) Macroscopic \ac{SEM} image showing the \ac{QTF} probe with the mounted \ac{TPP} 3D-printed exchangeable tip approaching the vertically oriented sample. 
(d) Closer \ac{SEM} view of the mounted \ac{TPP} 3D-printed exchangeable tip engaged with the sample surface. 
(e) High-magnification inset detailing the apex of the tip positioned directly over the region of interest.}
\label{fig6}
\end{figure}

The experimental setup for correlative \ac{AFM}, \ac{SEM}, and \ac{CL} imaging consists of a LiteScope \ac{SPM} module (NenoVision) mounted on the stage of an \ac{SEM} (TESCAN VEGA II). The system utilizes a custom dual-sample configuration, described here in a reference $0^{\circ}$ \ac{SEM} stage tilt position (Figure~\ref{fig6}a). In this configuration, the tip magazine carrying the prepared devices is mounted horizontally, and the sample of interest is mounted vertically. The seamless transition between the mounting regime and the measurement regime is facilitated by the \ac{SEM} stage rotation and tilting, combined with an optimized custom \ac{QTF} probe design (Figure~\ref{fig6}b). The \ac{QTF} is positioned horizontally, with the optical fiber attached to one of its prongs, ensuring unobstructed access to the tip magazine while preserving the \ac{QTF} probe's perpendicular orientation relative to the target sample. To ensure charge dissipation during \ac{SEM} operation, the probe is equipped with a grounded outer conductive layer. The complete fabrication workflow of the \ac{QTF} probe is detailed in Supplement 1 (Figure~S8). The mechanical quality factor of the finalized probes was typically in the low hundreds, with the best performers exceeding 1300.

In the measurement regime, the \ac{SEM} stage is tilted to $40^{\circ}$, providing an unobstructed view of the tip engaged with the sample (Figure~\ref{fig6}c--e). Consequently, the relative tilt angle between the electron beam and the sample surface is $50^{\circ}$. The apparatus can acquire up to three distinct signals for post-acquisition spatial correlation. The surface morphology is sensed by a \acf{SE} detector (\ac{SEM}), the 3D topography is derived from tip-sample interactions modulating the \ac{QTF} oscillations (\ac{AFM}), and the \ac{CL} distribution is mapped via the photons emitted from the sample upon electron beam irradiation. In the last case, the tip collects the emitted light, couples it into the optical fiber, and directs it to a \ac{PMT} (Hamamatsu). During scanning, the tip oscillates in a shear mode (amplitude regime, 90\,\% setpoint), while the electron beam is focused onto the surface with a defined spatial offset relative to the tip apex. The image formation is provided by scanning the sample via a piezo scanner relative to the stationary tip and electron beam.

\begin{figure}[htbp]
\centering\includegraphics{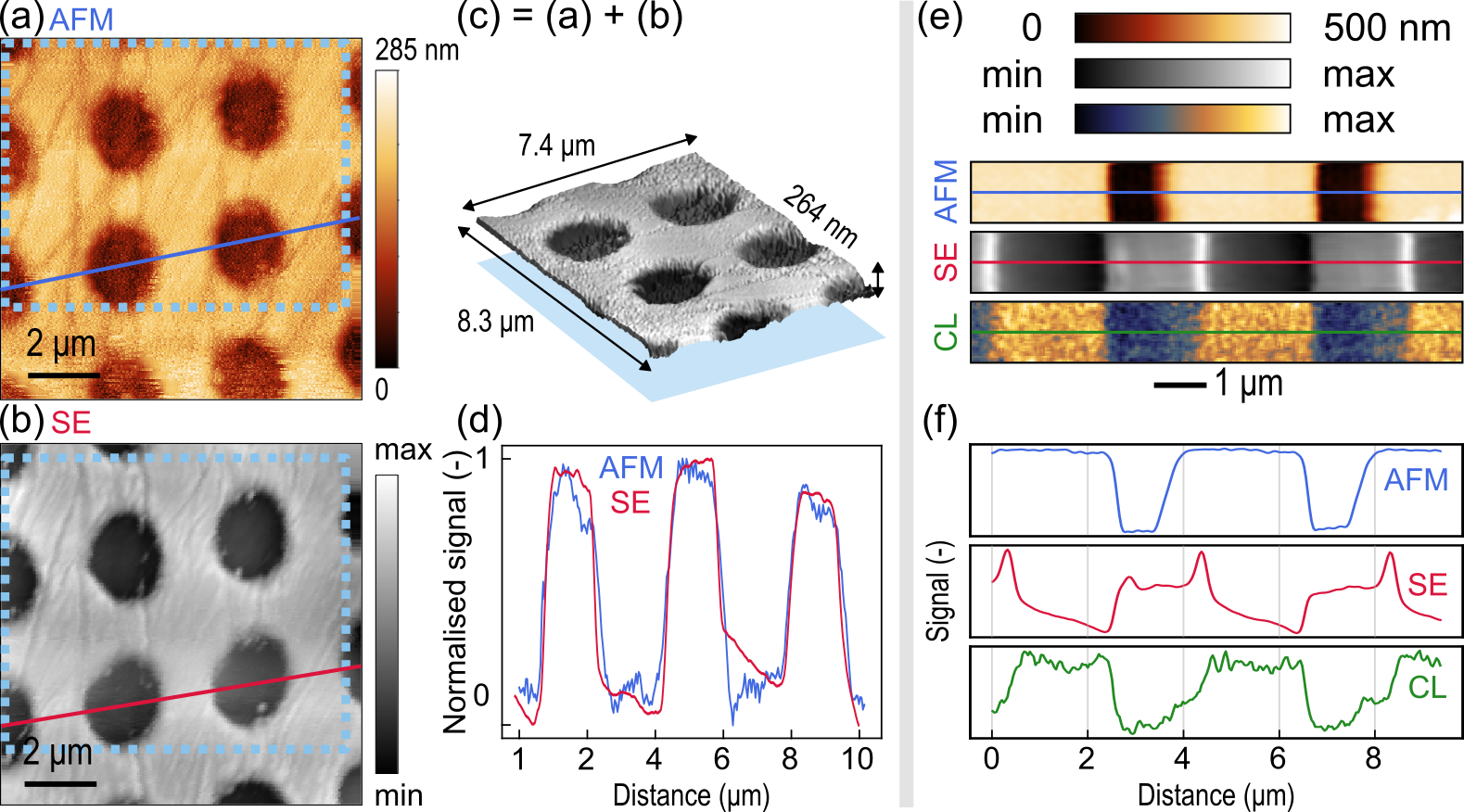}
\caption{Proof-of-concept correlative imaging utilizing the in situ mounted \ac{TPP} 3D-printed exchangeable \ac{SPM} tip. The figure demonstrates the mechanical stability of the modular platform and optical coupling capabilities during multimodal scanning probe microscopy.
(a) \ac{AFM} topography of the sample surface.
(b) Simultaneously acquired \ac{SE} image of the same region ($50^{\circ}$ sample tilt).
(c) 3D visualization of the correlated overlay, combining the \ac{AFM} topography and \ac{SE} signal from the region of interest indicated by the light blue dashed rectangles in (a) and (b).
(d) Normalized and shift-adjusted line profiles extracted along the blue and red solid lines in (a) and (b), respectively.
(e) Three-channel correlative imaging displaying simultaneously acquired \ac{AFM} topography (top), \ac{SE} (middle), and \ac{CL} signals (bottom).
(f) Line profiles extracted along the respective colored lines in (e), presented without shift adjustments.
All presented line profiles were acquired as an average across a 10-pixel-wide line.}
\label{fig7}
\end{figure}

The mechanical integrity of the modular platform was validated with dual-channel correlative measurements, simultaneously acquiring \ac{AFM} topography and \ac{SE} signals. A calibration grid (gold layer on a ruby substrate) featuring circular holes in the thin metal film with a step height of $200\,\mathrm{nm}$ was utilized as the sample. As shown in Figure~\ref{fig7}a,b, the platform successfully resolved the microstructured surface. The spatial correlation of the acquired images is visualized via a 3D overlay combining the two signals (Figure~\ref{fig7}c) and through normalized, shift-adjusted line profiles extracted from both data channels (Figure~\ref{fig7}d).

Following the mechanical validation, a three-channel measurement was conducted to verify the platform's optomechanical stability. A calibration grid consisting of line trenches in a $388\,\mathrm{nm}$ aluminum layer deposited on a Ce:YAP substrate served as the sample. Because Ce:YAP exhibits strong \ac{CL} emission upon electron excitation, the \ac{CL} image is expected to display alternating bright stripes corresponding to the exposed substrate, interrupted by dark regions where the aluminum layer blocks photon emission. The acquired three-channel dataset (Figure~\ref{fig7}e) demonstrates this periodic spatial modulation across all three signals, as highlighted by the raw, unadjusted line profiles in Figure~\ref{fig7}f. Compared to the dual-channel measurement, the \ac{CL} signal collection required higher electron beam currents, providing demanding operational conditions that induced localized tip and sample charging and image drift. Nevertheless, the successful simultaneous extraction of the \ac{CL} signal demonstrates the capability of the mounted tip to couple emitted light into the optical fiber.

By design, the modular socket creates an air gap between the optical fiber and the printed polymer component, replacing the direct fiber--polymer interface with separate fiber--air and air--polymer interfaces. This configuration adds extra Fresnel reflections and can lead to higher losses than in structures printed directly onto the fiber. The influence of varying the air gap was experimentally investigated in \cite{s3DPrintedFiber2020}, where a $5\,\upmu\mathrm{m}$ air gap reduced the measured transmittance from 69.35\,\% for direct contact to 66.5\,\%. Nevertheless, the present platform is intended for applications in which rapid modularity and consumable batch replacement are prioritized over maximum signal transmission. Furthermore, it is particularly well suited to devices that inherently incorporate an optical gap in their design, as proposed in Figure~\ref{fig1}c(ii)--(iv).

\section{Conclusions}

Here, we have presented a novel approach to optical fiber devices using fiber-decoupled fabrication in a horizontal orientation. This is facilitated by a \ac{TPP} 3D-printed modular platform that carries the device on its front and houses the optical fiber in a spring-like socket body. The advantages of fiber-decoupled fabrication are demonstrated through parallel post-processing, enabling batch treatment of devices using standard processes (such as curing, plasma etching, and metallization) while avoiding cumbersome optical fiber bundle management.

The concept is validated through an \ac{SPM} tip-device application, demonstrating proof-of-concept correlative two- and three-channel \ac{SE}, \ac{AFM}, and \ac{CL} measurements. Admittedly, state-of-the-art \ac{SPM} applications will require further application-specific optimization. However, the successful extraction of correlative \ac{AFM}, \ac{SEM}, and \ac{CL} data verifies the platform's optomechanical stability on the optical fiber, even under the high-stress conditions of shear-mode scanning.

Furthermore, this application demonstrates the opportunity for innovation in established fields enabled by fiber-decoupled fabrication in combination with plug-and-play functionality and self-alignment. We successfully implemented the planar substrate carrying the fabricated modular platform (i.e., exchangeable \ac{SPM} tips) as a tip magazine in an \ac{AFM}/\ac{SEM} setup, demonstrating the capability to mount and dismount devices in situ on the optical fiber. Additionally, we demonstrated a complementary manual mounting approach for benchtop environments.

Future work should focus on optimizing the \ac{SPM} application. Specifically, optimizing the design toward tapping-mode measurements will minimize lateral shear forces and mitigate tip wear. Furthermore, orienting the tip parallel to the printing laser will eliminate its inherent structural asymmetry. Additionally, demonstrating other proposed devices on the platform will showcase the broader versatility of this approach. \\

\noindent\textbf{Funding} This research was supported by Quantum materials for applications in sustainable technologies (QM4ST)--project No. CZ.02.01.01/00/22$\_$008/0004572 under OP JAK, call Excellent Research. L.Z. acknowledges financial support from the Brno Ph.D. Talent Scholarship, funded by the Brno City Municipality. 

\noindent\textbf{Acknowledgment}
CzechNanoLab Research Infrastructure (ID 90251), funded by MEYS CR, is gratefully acknowledged for the financial support of the measurements/sample fabrication. 

\noindent\textbf{Disclosures}
M.M. is an employee of IQS nano s.r.o., and Š.Č. was employed by IQS nano s.r.o. during the course of this research. The other authors declare no conflicts of interest. The authors used Gemini (3.1 Pro and 4 Argon) for grammar refinement and organizational suggestions. The tool was not used to generate scientific claims, references, experimental data, or conclusions. 

\noindent\textbf{Data Availability Statement}
The data that support the findings of this study are openly available in Zenodo at \url{https://doi.org/10.5281/zenodo.23208365}.

\noindent\textbf{Supplemental document}
See Supplement 1 for supporting content.

\singlespacing
\small{
    \bibliographystyle{unsrturl}
	\bibliography{literature} 
}

\clearpage

\rhead{Supporting Information}

\setcounter{section}{0}
\setcounter{figure}{0}
\setcounter{table}{0}
\setcounter{equation}{0}
\setcounter{page}{1}

\renewcommand{\thesection}{S\arabic{section}}
\renewcommand{\thefigure}{S\arabic{figure}}
\renewcommand{\thetable}{S\arabic{table}}
\renewcommand{\theequation}{S\arabic{equation}}
\renewcommand{\thepage}{S\arabic{page}}

\begin{center}
    \vspace*{1cm}
    {\LARGE \textbf{Supporting Information for:}}\\[0.5cm]
    {\Large \textbf{A plug-and-play platform for 3D-printed devices on optical fibers}}\\[1cm]

    {\large 
    Lukáš Zezulka$^{1,2}$, Zdeněk Nováček$^{1,2,\dagger,*}$, Martin Konečný$^{1,2}$, \\
    Milan Matějka$^{3,4}$, Štěpán Černý$^{4}$, Katarína Rovenská$^{1,2}$, \\
    Jiří Spousta$^{1,2}$, and Tomáš Šikola$^{1,2}$
    }\\[1cm]

    {\small 
    $^{1}$Brno University of Technology, Faculty of Mechanical Engineering, Institute of Physical Engineering, Technická 2, 61669 Brno, Czech Republic\\[0.2cm]
    $^{2}$Brno University of Technology, Central European Institute of Technology, Purkyňova 656/123, 61200 Brno, Czech Republic\\[0.2cm]
    $^{3}$Institute of Scientific Instruments of the Czech Academy of Sciences, Královopolská 147, 61264 Brno, Czech Republic\\[0.2cm]
    $^{4}$IQS nano s.r.o., Hlavní 130, 25068, Husinec, Czech Republic\\[0.5cm]
    
    $^{\dagger}$These authors contributed equally to this work.\\
    $^{*}$Corresponding author. E-mail: novacek@fme.vutbr.cz
    }
\end{center}
\vspace{1cm}

\begin{abstract}
This document provides and documents the supplementary information to the main manuscript. Included are individual top- and side-view scanning electron microscopy images of the individual characterized two-photon polymerization 3D-printed tips, details on the influence of gas mixtures during the plasma etching protocol, a step-by-step description of the quartz tuning fork probe assembly, and descriptive text for Visualization 1.
\end{abstract}

\section{Morphological Characterization of Printed Batches}
This section groups the top-view and side-view images for each of the three batches sequentially, allowing the reader to cross-reference the visual evidence with the measured data easily.

\subsection{Batch 1 Characterization}

\begin{figure}[htbp]
\includegraphics[width=\linewidth]{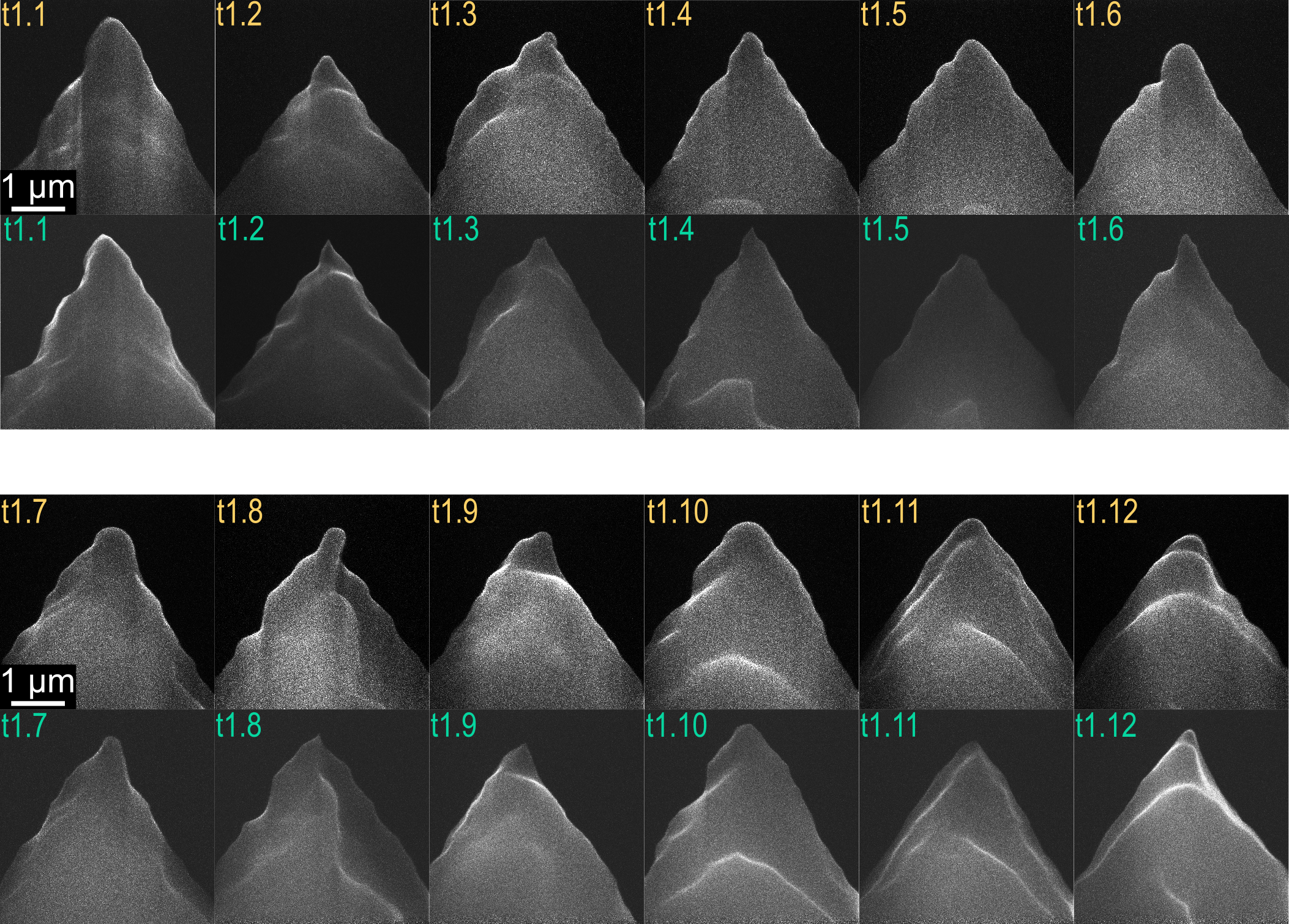}
\caption{Top-view morphological characterization of the \ac{TPP} 3D-printed tips from the first fabrication batch. The \ac{SEM} images display all 12 printed samples from this batch (labeled t1.1 to t1.12), comparing the apex geometries in the as-printed state (top rows, yellow labels) versus the etched state (bottom rows, green labels). To strictly avoid any additional influence on the tip shape and dimensions, all samples were imaged in their pristine state without applying conductive coatings.}
\label{SIfig:b1t}
\end{figure}

\begin{figure}[htbp]
\centering\includegraphics{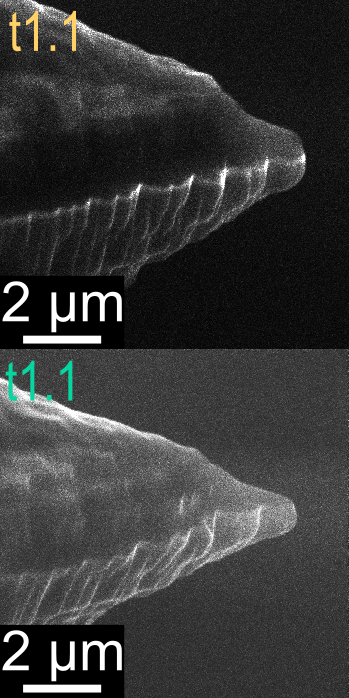}
\caption{Side-view morphological characterization of a \ac{TPP} 3D-printed tip from the first fabrication batch. The \ac{SEM} images illustrate the vertical profile of tip t1.1 in the as-printed (left, yellow label) and etched (right, green label) conditions. Similar to the top-view analysis, the sample was imaged in its pristine state without any conductive coatings. While this fabrication batch consists of 12 successfully printed samples, only a single tip (t1.1) was fully imaged from the side due to the gradual development of the imaging methodology at this early stage of the study.
}
\label{SIfig:b1s}
\end{figure}

\clearpage
\subsection{Batch 2 Characterization}

\begin{figure}[htbp]
\centering\includegraphics[width=\linewidth]{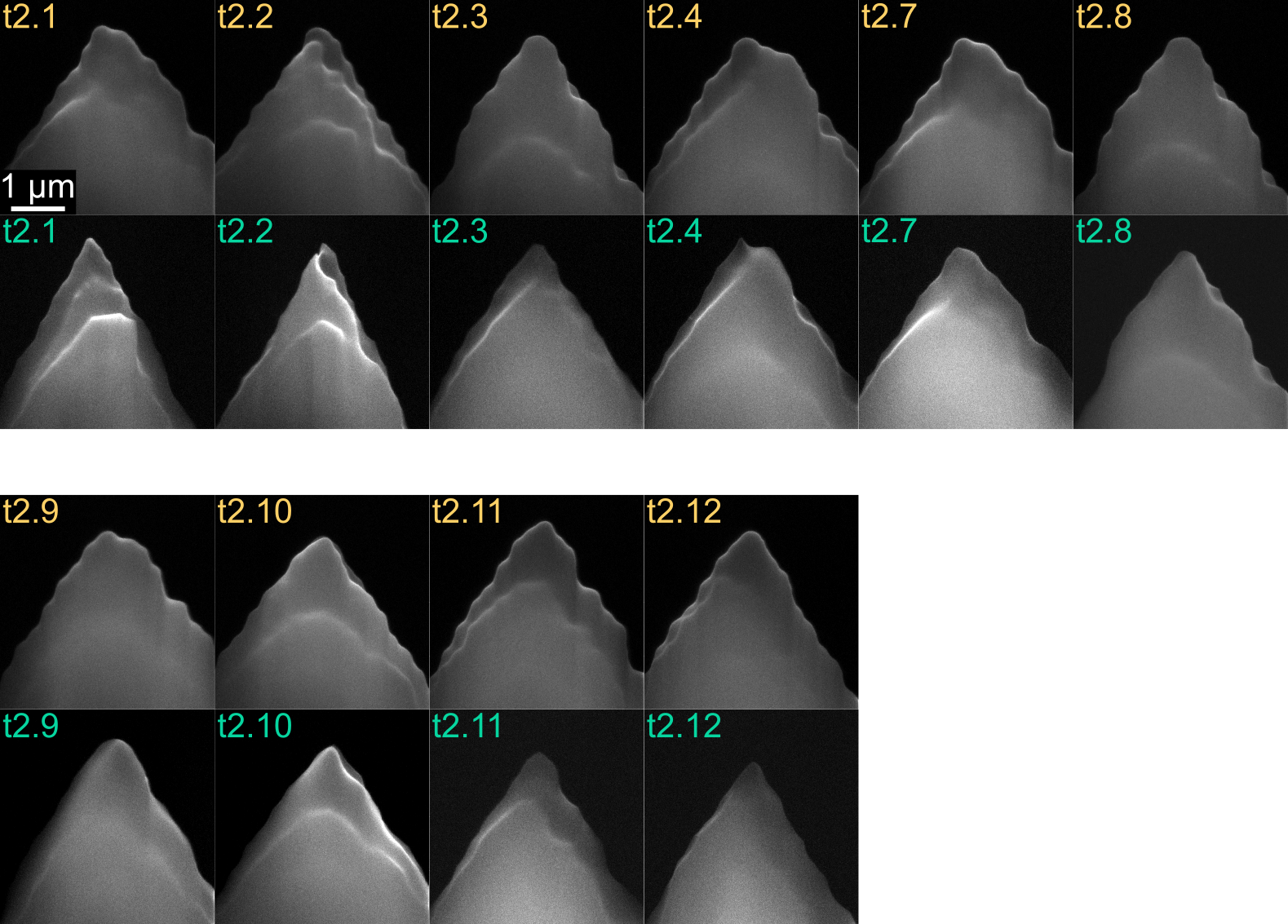}
\caption{Top-view morphological characterization of the \ac{TPP} 3D-printed tips from the second fabrication batch. The \ac{SEM} images display 10 samples from this batch (labeled t2.1 to t2.4 and t2.7 to t2.12), comparing the apex geometries in the as-printed state (top rows, yellow labels) versus the etched state (bottom rows, green labels). Although a nominal batch consists of 12 printed samples, two tips (t2.5 and t2.6) are missing here because they were skipped in the printing queue due to an error. To strictly avoid any additional influence on the tip shape and dimensions, all samples were imaged in their pristine state without applying conductive coatings.
}
\label{SIfig:b2t}
\end{figure}

\begin{figure}[htbp]
\centering\includegraphics{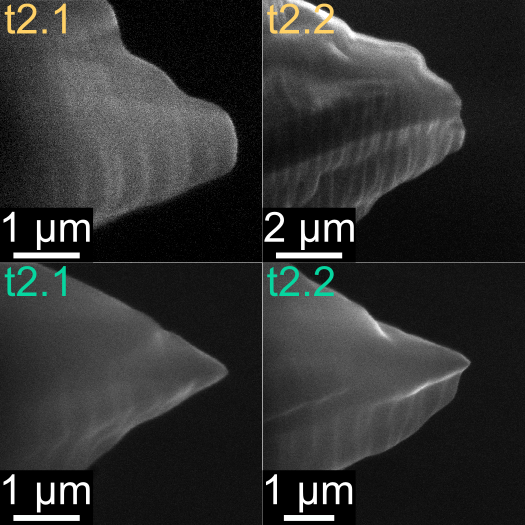}
\caption{Side-view morphological characterization of \ac{TPP} 3D-printed tips from the second fabrication batch. The \ac{SEM} images illustrate the vertical profiles of tips t2.1 and t2.2 in the as-printed (top row, yellow labels) and etched (bottom row, green labels) conditions. Similar to the top-view analysis, the samples were imaged in their pristine state without any conductive coatings. While 10 samples were successfully printed in this batch, only two tips were fully inspected from the side due to the gradual development of the imaging methodology at this stage of the study.
}
\label{SIfig:b2s}
\end{figure}

\clearpage
\subsection{Batch 3 Characterization}

\begin{figure}[htbp]
\centering\includegraphics[width=\linewidth]{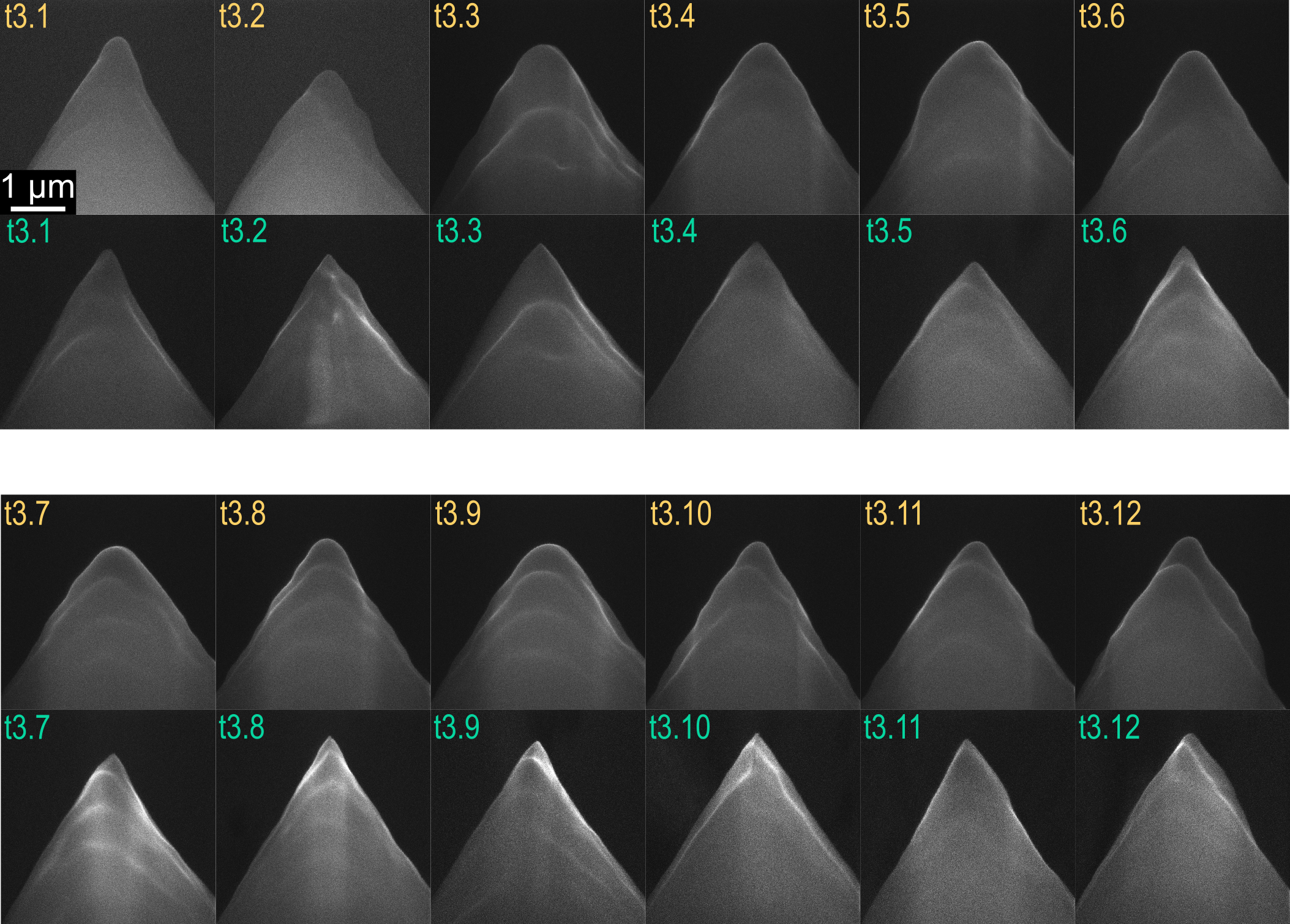}
\caption{Top-view morphological characterization of the \ac{TPP} 3D-printed tips from the third fabrication batch. The \ac{SEM} images display all 12 printed samples from this batch (labeled t3.1 to t3.12), comparing the apex geometries in the as-printed state (top rows, yellow labels) versus the etched state (bottom rows, green labels). To strictly avoid any additional influence on the tip shape and dimensions, all samples were imaged in their pristine state without applying conductive coatings.}
\label{SIfig:b3t}
\end{figure}

\begin{figure}[htbp]
\centering\includegraphics[width=\linewidth]{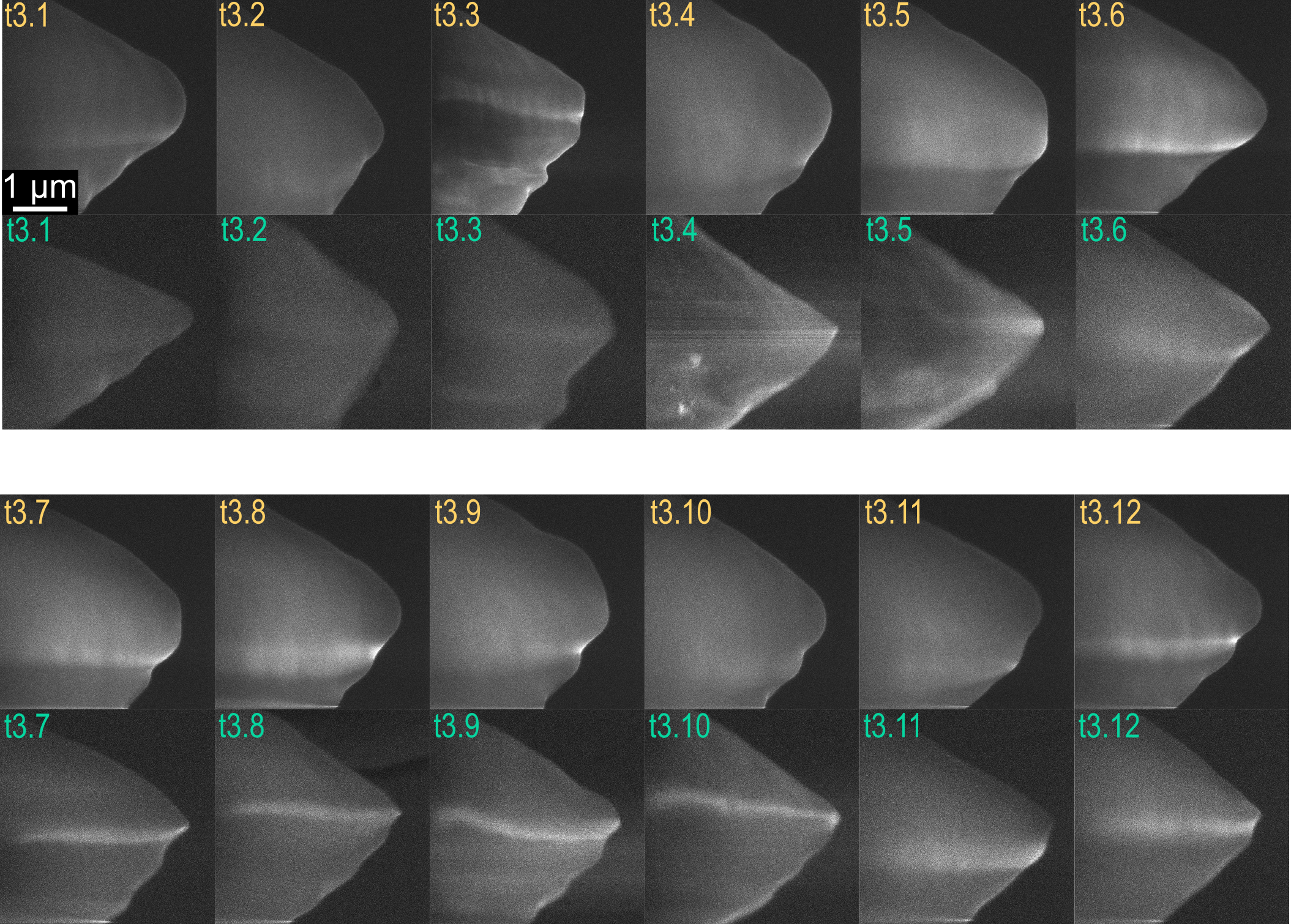}
\caption{Side-view morphological characterization of the \ac{TPP} 3D-printed tips from the third fabrication batch. The \ac{SEM} images illustrate the vertical profiles of all 12 printed samples from this batch (labeled t3.1 to t3.12) in the as-printed (top rows, yellow labels) and etched (bottom rows, green labels) conditions. Similar to the top-view analysis, the samples were imaged in their pristine state without any conductive coatings. Unlike the earlier batches, the imaging methodology was fully developed by this stage of the study, allowing every tip in this batch to be successfully inspected from the side.}
\label{SIfig:b3s}
\end{figure}

\clearpage
\section{Etching Protocol Optimization}
This section justifies the specific gas mixture utilized in the final experimental methodology.

\begin{figure}[htbp]
\centering\includegraphics{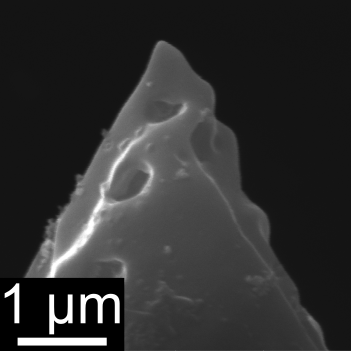}
\caption{Top-view \ac{SEM} image of a single \ac{TPP} 3D-printed tip demonstrating the bulging effect observed during the etching protocol optimization. The image illustrates the severe structural degradation and bulging resulting from etching the polymer exclusively with O$_2$ plasma ($30\,\mathrm{sccm}$) for 5 minutes. This adverse morphological change highlights the necessity of the optimized etching protocol utilized in the main fabrication batches, which instead employs a 1:1 gas mixture of O$_2$ and SF$_6$ to achieve uniform tip sharpening without such detrimental artifacts.}
\label{SIfig:O2etch}
\end{figure}

\clearpage
\section{Quartz tuning fork probe fabrication}

\begin{figure}[htbp]
\centering\includegraphics[width=\linewidth]{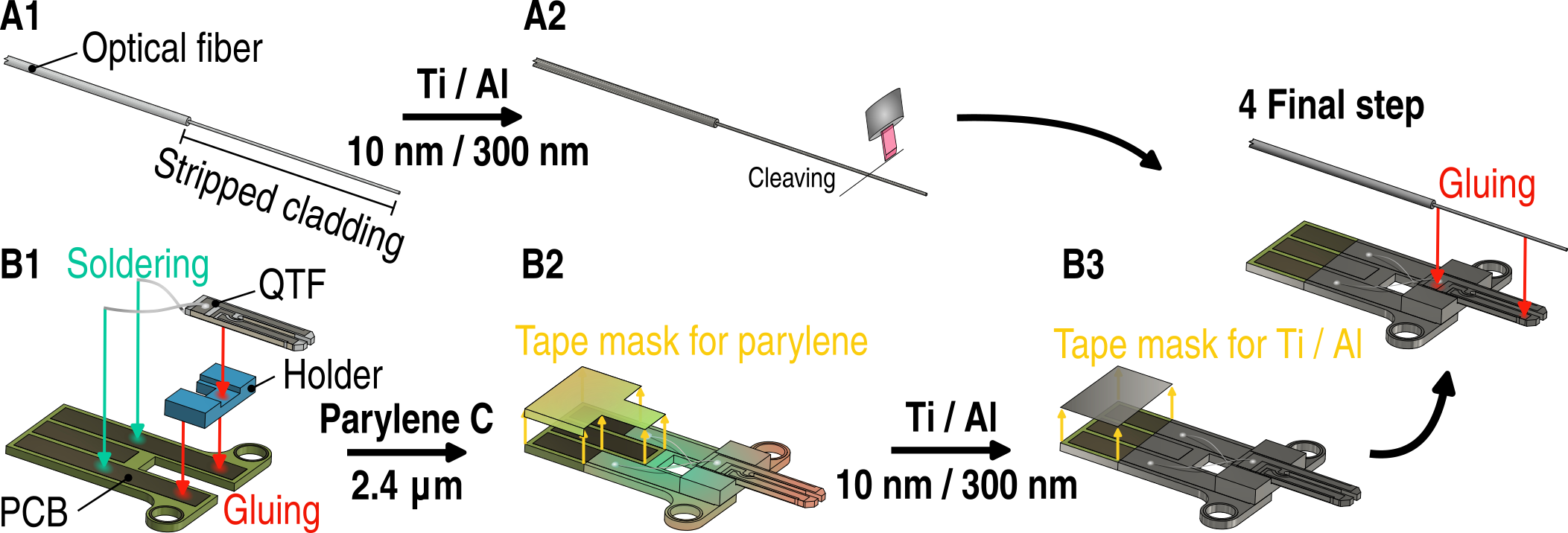}
\caption{Visualization of the \ac{QTF} probe fabrication process.
    A1 Optical fiber with partially stripped cladding.
    A2 Cleaving of the optical fiber after metal sputtering to create a clean facet.
    B1 Assembly and contacting of the \ac{PCB}, \ac{QTF}, and its holder.
    B2 Parylene C deposition over a tape mask.
    B3 Metal sputtering over a new tape mask.
    4 Final step of gluing the metallized optical fiber to the metallized probe body.
    }
\label{SIfig:QTFfab}
\end{figure}

The complete fabrication procedure of the probe is illustrated in Figure~\ref{SIfig:QTFfab}. The process starts with an FG105UCA multimode optical fiber (Thorlabs) with a core diameter of $105\,\upmu\mathrm{m}$ and a cladding diameter of $125\,\upmu\mathrm{m}$ \cite{ThorlabsFG105UCA}, from which the front section of the cladding is stripped (step A1). The stripped segment is then coated by sputtering $10\,\mathrm{nm}$ of Ti, followed by $300\,\mathrm{nm}$ of Al, using a broad-beam ion source while rotating the fiber (step A2). 

Next, the probe body is assembled by gluing the \ac{QTF} into a 3D-printed holder, which is then attached to the \ac{PCB}. After mechanical assembly, the \ac{QTF} is electrically connected to the contact pads on the \ac{PCB} via two wires (step B1). The contact pads are then partially protected with Kapton tape, and a $2.4\,\upmu\mathrm{m}$ Parylene~C layer is deposited over the entire probe body using an SCS Parylene Deposition System (step B2). This dielectric layer insulates the electrical contacts of the \ac{PCB} and \ac{QTF}, preventing short-circuiting during subsequent metallization. Afterward, a bilayer of $10\,\mathrm{nm}$ Ti and $300\,\mathrm{nm}$ Al is deposited on the rotating probe body with a second Kapton tape mask applied (step B3) using the \ac{IBSD} system. This metallic coating establishes a conductive, grounded outer surface that prevents charging of the tip and probe inside the \ac{SEM}. Importantly, the two tape masks are designed so that the outer metal layer self-contacts with the central \ac{PCB} pin, which serves as the grounding point.

Finally, the pre-coated optical fiber (steps A1–A2) is glued onto the tuning fork. Adhesive is applied sparingly at the front of the prong and more generously at the back, reducing disruption of the resonator whilst maintaining mechanical stability (step 4).

\section{Visualization 1}
Visualization 1 demonstrates the in situ \ac{SEM} mounting process of the \ac{TPP} 3D-printed exchangeable \ac{SPM} tip onto an optical fiber. While this recording captures a different experimental iteration than the specific sequence shown in Figure~5(b1–b5) of the main manuscript, it illustrates the same workflow. The video shows the optical fiber alignment using the guiding block, its insertion into the tip's spring-like socket through the guiding cone, and the final detachment of the tip from its supporting struts. 

Furthermore, the video includes an earlier, now-retired force-mounting step utilizing a second support block; because this step is no longer applied, it is not described in the main text. Please note that the video playback has been accelerated, and static periods corresponding to \ac{SEM} image acquisition have been edited out for brevity.

For convenience, it can also be viewed online at \url{https://youtu.be/Pwhci0SOXpE}.

\end{document}